\documentstyle[12pt]{article}
\begin{document}
\begin{center}
{\Large{\bf
O(d,d)-invariance in inhomogeneous string cosmologies with perfect fluid
}}
\vspace*{10mm} \\
Jacques Demaret\footnote{E-mail: demaret@astro.ulg.ac.be} and 
Elisa Di Pietro\footnote{E-mail: dipietro@astro.ulg.ac.be} 
\vspace*{5mm} \\
Institute of Astrophysics and Geophysics \vspace*{3mm} \\
Group of Theoretical Cosmology \vspace*{3mm} \\
University of Li\`ege \vspace*{3mm} \\
B-4000 LIEGE-BELGIUM \vspace*{23mm} \\

\begin{abstract}
\noindent
In the first part of the present paper, we show that $O(d,d)$-invariance
usually known in a homogeneous cosmological background written
in terms of proper time can be extended to backgrounds depending on one or
several coordinates (which may be any space-like or time-like coordinate(s)).
In all
cases, the presence of a perfect fluid is taken into account and the 
equivalent duality transformation in Einstein frame is explicitly given. In 
the second part, we present several concrete applications to some
four-dimensional metrics, including inhomogeneous ones, which illustrate the
 different duality transformations
discussed in the first part. Note that most of the dual solutions given here
do not seem to be known in the literature. 
\end{abstract}
\end{center}
\hspace*{5mm} \vspace*{3mm} \\
\underline{Key words:} superstring cosmology - duality - exact solutions -
pre-Big Bang.
\newpage
\section{Introduction.}
As it is well known, the standard cosmological model describes remarkably 
well most of the present Universe's properties: extragalactic sources 
redshift, light elements nucleosynthesis, the cosmic microwave background 
at 2.735 K, ... In spite of these successes, there are some problems that 
this model cannot solve: initial singularity, horizon problem, isotropy, 
flatness and dimensionality of the present Universe, ... Much work has been 
done in order to solve these problems, in particular in the framework of 
inflationary models (see e.g. \cite{infl1} and \cite{infl2} and the 
references therein), but these models have in general to be fine tuned in 
order to give expected results.
\vspace*{2mm} \\
Some years ago appeared string cosmology \cite{string}. We do not consider
here string theory as issued from particle physics and quantum field theory
in its full scope, but we restrict rather to its low-energy approximation,
i.e. string effective theory viewed as an alternative theory to general
relativity. String cosmology is characterized by the corresponding field
equations which generalize Einstein's field equations. This theory contains
a metric tensor (as in general relativity), a scalar field, called
{\it dilatonic field}, and
a rank-three antisymmetric tensor field, called {\it axionic field}. String 
theory has motivated the study of cosmo\-lo\-gi\-cal models because of its 
most remarkable property of symmetry called {\it T-duality} ({it scale factor
duality} in the framework of effective theory) \cite{vene1}.
This symmetry allows, from a string theory solution, to construct a second 
solution, called {\it dual solution}, so that in string theory, the first 
solution is used to describe the ``{\it post-Big Bang phase}'', i.e. the 
Universe's evolution between the initial singularity until today, whereas the
dual solution describes an eventual ``{\it pre-Big Bang phase}'' \cite{odd}, 
i.e. the Universe as it could be before the Big Bang. Though there is no 
satisfying model explaining the transition between pre- and post-Big Bang, 
many cosmologists think that duality symmetry and in particular the pre-Big 
Bang phase could solve the singularity problem.
\vspace*{2mm} \\
Duality symmetry associates with the present Universe (decelerated expanding 
Universe) an accelerated expanding Universe \cite{kar}, so that the pre-Big 
Bang solution appears naturally as an inflationary phase and can in
principle solve the isotropy and flatness problems. Furthermore, string 
theory is initially a 10 or 26-dimensional theory, depending on whether 
heterotic or bosonic strings are considered, but at low energy, this theory 
allows the compactification of some of these dimensions and the expansion of 
the others. So one can hope to explain the present dimensionality of the 
Universe by the compactification of the extra dimensions of string theory.
\vspace*{2mm} \\
As one can see, one can expect a lot from string theory and in particular
from the pre-Big Bang phase that appears naturally in this theory. Most of 
the work done so far on duality has been based on Meissner {\it et al.}'s 
paper \cite{vene1}. The authors present a duality transformation that 
can be applied to a homogeneous solution written in proper time only. Our
aim is to show that duality symmetry can be applied to every kind of 
cosmological string solution. We shall show more particularly how to
generalize Meissner {\it et al.}'s transformation so that it can be
applied to:
\begin{itemize}
\item a homogeneous solution written in terms of a time coordinate that can
be different from proper time. Indeed sometimes field equations cannot be
solved in terms of proper time so that cosmological solutions are known in
terms of another
time coordinate and the integration allowing the passage from this time
coordinate to proper time is not always possible. So if we want to build the
dual of such a solution, it is necessary to extend Meissner {\it et al.}'s 
transformation to any time-like coordinate. 
\item a solution depending on one space coordinate. This could be the case
for a spherically symmetric black hole solution. As one can hope that there
exists a possibility of avoiding the Big Bang singularity in the framework of
the pre-Big Bang scenario, one can imagine to overpass the black hole
singularity using a similar mechanism.
\item a more general solution that can depend on several coordinates. Indeed,
we know that near the cosmological singularity, our Universe is neither 
isotropic nor homogeneous. Accordingly more general models have to be used to study
the primordial Universe.
\end{itemize}
In all the preceding cases, we have supposed that space-time is filled with a
perfect fluid. Moreover, as some authors prefer to consider Einstein frame
as the physical frame, we have also developed the duality transformation
directly in Einstein frame.
\vspace*{2mm} \\
The present paper is organized as follows: in the first section, we establish 
all the theoretical duality transformations in string as well as in Einstein
frames. First we write explicitly the duality transformation in the two 
frames in the case of fields depending on one coordinate which may be
different from a time-like coordinate. But when this coordinate is time-like,
our duality transformation is valid for any time coordinate used: proper 
time, logarithmic time, conformal time, ... Then we extend our results to 
the most general case of fields depending on several coordinates. For each
case, duality transformation of the energy-momentum tensor is explicitly
displayed. 
\vspace*{2mm} \\
In the second section, we apply our theoretical duality transformation to
some concrete examples. Note that most of the dual solutions presented here 
are not known in the literature. Each of these examples has been chosen to 
illustrate a specific aspect of duality transformation. Duality is performed
on solutions written in terms of proper time and of logarithmic time,
corresponding to  four-dimensional homogeneous and inhomogeneous metrics. Most
 of these solutions are given in presence of a perfect
fluid as well in both frames. In the last section, we briefly comment on the
possibility of analyzing geometrical and physical properties of the dual
solution obtained in section 3 in view of their eventual application to the
pre-Big Bang scenario.
\vspace*{2mm} \\
Note that the explicit form of all the relations presented in this paper 
(duality transformations, solutions, dual solutions, ...) has been checked by
using symbolic programming (Reduce, Excalc, Mathematica, ...).
   
\section{O(d,d)-invariance of string theory.}
\subsection{Duality when all fields depend on one coordinate only
(space-like or time-like).}
\underline{i. In the string frame.}  \vspace*{3mm} \\
At low energy, the tree-level effective action for closed superstring theory 
(in its bosonic sector) can be written as
\begin{equation}
S_{eff} = \frac{1}{2 \kappa^2_D}\,\int{ \left( 
e^{-\phi}\,\left[ R + \left[ \nabla \phi \right]^2  
- \frac{1}{12} \,H_{\alpha\beta\delta}\,H^{\alpha\beta\delta} \right] 
+ L_m \right)
\sqrt{\mid det\,g \mid }\,\,d^D x}
\label{action1}
\end{equation}
where 
\begin{center}
\begin{tabular}{ll}
$D$ & is the space-time dimension, \\
$\kappa^2_D$ & is a parameter connected with the fundamental string
length, \\
$det\,g$ & is the determinant of the metric tensor $g_{\alpha\beta}$, \\
$R$ & is the curvature scalar, \\
$\phi$ & is the massless dilatonic scalar field, \\
$H_{\alpha\beta\delta}$ & is the completely antisymmetric tensor field 
strength defined  \\
& by $H = d B$, where $B$ is a rank-two antisymmetric tensor, \\
$L_m$ & is the Lagrangian for the matter (perfect fluid), \\
$\left[ \nabla \phi \right]^2$ & stands for $g^{\alpha\beta} \nabla_\alpha 
\phi \nabla_\beta \phi$, where $\nabla_\alpha$ is the covariant derivative \\
& with respect to $x^\alpha$.
\end{tabular} 
\end{center}
By varying the action with respect to $g_{\alpha\beta}$, $\phi$ and 
$B_{\alpha\beta}$, we can find respectively the following field 
equations\footnote{Greek indices run from $0$ to $D-1$.} \cite{copel1} 
\begin{equation}
R^{\alpha}_{\hspace{2mm}\beta} - \frac{1}{4} H^{\alpha\mu\nu}\,H_{\beta\mu\nu} 
+ g^{\alpha\delta} \nabla_\delta\nabla_\beta\,\phi 
= \kappa^2_D \,e^\phi\,T^{(m)\alpha}_{\hspace{7mm}\beta} 
\label{eqncordes}
\end{equation}
\begin{equation}
R + 2\,\Box\,\phi - \left[ \nabla \phi \right]^2 
- \frac{1}{12} H^2 = 0 
\label{eqncordes2}
\end{equation}
\begin{equation}
\nabla_\mu\, \left( e^{-\phi}\,H^{\mu\alpha\beta}\right) = 0
\label{eqncordes3}
\end{equation}
where $T^{(m)}_{\alpha\beta}$ is the energy-momentum tensor derived from the
Lagrangian $L_m$ and $\Box$ stands for the dalembertian operator.
\vspace*{3mm} \\
As we shall see, the action (\ref {action1}) is invariant under a symmetry 
transformation called ``{\it O(d,d)-invariance}'' (T-duality) where $d$ is the 
number of coordinates the metric and other fields do not depend on. This 
invariance is well known for a particular homogeneous cosmological background 
without matter, i.e. for a metric $g_{\alpha\beta}$, a dilaton $\phi$ and a 
potential $B_{\alpha\beta}$ depending on time only and for
$T^{(m)}_{\alpha\beta} = 0$ (see e.g. \cite{vene1}). 
We shall first extend preceding work \cite{vene1} to the case of fields 
depending on one coordinate only (which may be different from the time-like
coordinate) in the presence of matter. Then, we shall show how the duality 
transformation can be generalized when all fields ($g_{\alpha\beta}$, $\phi$, 
$T^{(m)}_{\alpha\beta}$ and $B_{\alpha\beta}$) depend on several coordinates.
\vspace*{3mm} \\
Let us first consider the case of $g_{\alpha\beta}$, $\phi$, 
$T^{(m)}_{\alpha\beta}$ and 
$B_{\alpha\beta}$ depending on one coordinate only that can be eventually
different 
from the time-like coordinate. For simplicity, we shall note the coordinates 
$x^0, ..., x^d$ where $d = D - 1$ and order them so that we can say
that the fields depend on $x^0$ and do not depend on $x^1$, ..., $x^d$. Note 
that $0$-index does not refer necessary to the usual time-like coordinate but 
to the coordinate the fields depend on, i.e. $x^0$. We shall also, as in 
\cite{vene1}, introduce the following assumptions 
\begin{equation}
g_{0i} = 0 \hspace{20mm} B_{0i} = 0
\label{hyp1}
\end{equation}
with (i = 1, ..., d), so we can write 
\begin{equation}
g_{\alpha\beta} = \left(
\begin{tabular}{cc}
$g_{00}$ & $0$ \\
$0$           & $G$
\end{tabular}
\right)  
\hspace{20mm} 
B_{\alpha\beta} = \left(
\begin{tabular}{cc}
$0$ & $0$ \\
$0$ & $B$
\end{tabular}
\right)
\label{defgandb}
\end{equation}
where $B$ and $G$ are $d \times d$ matrices and $g_{00}$ is the metric's 
component relative to the $x^0$ coordinate; $B_{ij}$, $G_{ij}$ ($i, j =
1,...,d$) and $g_{00}$ are functions of $x^0$. 
\vspace*{3mm} \\
Defining respectively the ``{\it shifted dilaton}'' $\Phi$ and the $d \times 
d$ matrix $M$
\begin{equation}
\Phi = \phi - ln \sqrt{\mid det\,g \mid} \hspace{19mm}
\label{defphi}
\end{equation}
\begin{equation}
M = \left( 
\begin{tabular}{cc}
$G^{-1}$   & $ - G^{-1} B$ \\
$B G^{-1}$ & $ G - B G^{-1} B$
\end{tabular}
\right)
\label{defm}
\end{equation}
we can rewrite the effective action (\ref {action1}) under the form
$$
S_{eff} = \frac{1}{2 \kappa^2_D} \,
\int{ e^{- \Phi}}\,\left( g^{00} \left[ \frac{1}{8} 
Tr \left[ \partial_0 M \eta \partial_0 M \eta \right] 
+ \left[\partial_0 \Phi \right]^2
\right]   \right. \hspace{30mm}
$$
\begin{equation}
\left.\hspace{40mm}
- \partial^2_0 g^{00} + \frac{1}{4}\,g_{00} \left[g^{00}\right]^2 
+ e^\phi \,L_m \right)\, d^D x
\label{action2}
\end{equation}
where $\eta = \left(
\begin{tabular}{cc}
$0$ & $I$ \\
$I$ & $0$
\end{tabular}
\right)$ is an element of the O(d,d)-group and $I$ stands for the 
$d \times d$ unit matrix\footnote{ 
When $x^0$ is the proper time, we are back to
the hypothesis used by Meissner {\it et al.} in \cite{vene1}. It is easy to 
see that in this case, our action (\ref {action2}) is the same as theirs.}. 
\vspace*{3mm} \\
It is easy to see that the part of action (\ref {action2}) which does not 
include matter is invariant under  
\cite{vene1} 
\begin{itemize}
\item the following transformation on $M$:
$M \rightarrow \bar{M} = \Omega\, M \Omega^T$ \vspace*{3mm} \\
where $\Omega$ is an element of the O(d,d)-group, such as  
$\Omega^T \eta \Omega = \eta$. Taking as usually $\Omega \equiv \eta$, we can 
write
\begin{equation}
\bar{M} = \left(
\begin{tabular}{cc}
$\bar{G}^{-1}$    & $- \,\bar{G}^{-1}\,\bar{B}$ \\
$\bar{B}\,\bar{G}^{-1}$ & $\bar{G} - \bar{B}\,\bar{G}^{-1}\,\bar{B}$
\end{tabular}
\right)
= \left(
\begin{tabular}{cc}
$G - B\,G^{-1}\,B $ & $ B\,G^{-1} $ \\
$- \,G^{-1}\,B $      & $ G^{-1} $ 
\end{tabular}
\right)
\label{mbarre}
\end{equation}
So equation (\ref {mbarre}) gives the duality transformation for $G$ and $B$
\begin{equation}
\left\{
\begin{tabular}{l}
$G \rightarrow \bar{G} = \left( G\, - B\,G^{-1}\,B \right)^{-1}$ \\
$B \rightarrow \bar{B} = -\,G^{-1}\,B\,
\left( G\, - B\,G^{-1}\,B \right)^{-1} $ \\
\end{tabular}
\right.
\label{transfo1}
\end{equation}
\item the following transformation on $\Phi$:
$\Phi \rightarrow \bar{\Phi} = \Phi$ \vspace*{3mm} \\
Using equation (\ref {defphi}), we can rewrite this transformation in terms 
of $\phi$ and $\bar{\phi}$: 
\begin{equation}
\phi \rightarrow \bar{\phi} = \phi + \frac{1}{2} \,ln
\left( \frac{det\,\bar{G}}{det\,G}
\right)
\label{transfo2}
\end{equation}
\item the following transformation on $g_{00}$:
\begin{equation}
g_{00} \rightarrow \bar{g}_{00} = g_{00} 
\label{transfo22}
\end{equation} 
\end{itemize}
\noindent
The transformations (\ref {transfo1})-(\ref {transfo22}) can be compared
with Busher's transformations (\cite{busher}, \cite{darkside}) obtained when 
hypothesis (\ref {hyp1}) is not realized and in absence of matter and 
potential $B$. Busher has shown that, starting from a solution with
$B=0$ and $g_{0i}\not=0$, a torsion potential $B$ is generated after duality.
\vspace*{3mm} \\
It remains now to see how the energy-momentum tensor must transform if we 
want that the complete action (\ref {action2}) stays invariant under duality. 
If we introduce transformations (\ref {transfo1})-(\ref {transfo22}) into
equation (\ref {eqncordes}) without making any change in the energy-momentum
tensor\footnote{In this paper, we shall always consider a diagonal 
energy-momentum tensor $T^{(m)\alpha}_{\hspace{8mm}\beta}$.}, we transform 
equation (\ref {eqncordes}) in
\begin{equation}
R^{\alpha}_{\hspace{2mm}\beta} - \frac{1}{4} H^{\alpha\mu\nu}\,H_{\beta\mu\nu} 
+ g^{\alpha\delta} \nabla_\delta\nabla_\beta\,\phi 
= - \kappa^2_D\,e^\phi\,\sqrt{\frac{det\,\bar{G}}{det\,G}}\,
T^{(m)\alpha}_{\hspace{7mm}\beta} 
\end{equation}
if $\alpha \not= 0$, and in
\begin{equation}
R^{\alpha}_{\hspace{2mm}\beta} - \frac{1}{4} H^{\alpha\mu\nu}\,H_{\beta\mu\nu} 
+ g^{\alpha\delta} \nabla_\delta\nabla_\beta\,\phi 
= \kappa^2_D\,e^\phi\,\sqrt{\frac{det\,\bar{G}}{det\,G}}\,
T^{(m)\alpha}_{\hspace{7mm}\beta} 
\end{equation}
if $\alpha = 0$, where it is important to keep in mind that 
$0$-index corresponds to the coordinate the metric and other fields depend on, so 
that $T^{(m)\,0}_{\hspace{7mm}0}$ is not necessary the energy density. 
\vspace*{3mm} \\
Thus, we can say that:
\begin{itemize}
\item if $\alpha \not= 0$, $T^{(m)\alpha}_{\hspace{7mm}\beta}$ must 
transform as follows
\begin{equation}
T^{(m)\alpha}_{\hspace{7mm}\beta} \rightarrow 
\bar{T}^{(m)\alpha}_{\hspace{7mm}\beta} = 
- \left[\frac{det\,\bar{G}}{det\,G} \right]^{-1/2}\,
T^{(m)\alpha}_{\hspace{7mm}\beta} 
\label{transfo3}
\end{equation}
\item if $\alpha = 0$, $T^{(m)\alpha}_{\vspace{7mm}\beta}$ must transform
as follows
\begin{equation}
T^{(m)\alpha}_{\hspace{7mm}\beta} \rightarrow 
\bar{T}^{(m)\alpha}_{\hspace{7mm}\beta} = 
\left[\frac{det\,\bar{G}}{det\,G} \right]^{-1/2}\,
T^{(m)\alpha}_{\hspace{7mm}\beta} 
\label{transfo4}
\end{equation}
\end{itemize}
In conclusion, the duality transformation in string frame is given by
the relations
(\ref {transfo1})-(\ref {transfo22}), (\ref {transfo3})-(\ref {transfo4}). 
We shall later on present  a few concrete examples of application of this
transformation for a few four-dimensional space-times.
\vspace*{3mm} \\
\underline{ii. In the Einstein frame.} \vspace*{3mm} \\
It is sometimes easier to solve the field equations when they are written 
in Einstein frame, so that, in some cases, it is more convenient to work in 
Einstein frame. We shall see that starting from a solution written in the 
Einstein frame, it is possible to obtain directly the dual solution in the
same frame\footnote{Of course, it is possible, starting from a solution 
given in Einstein frame, to write the corresponding solution in string 
frame, via the conformal transformation (\ref {conforme}) and to find in this
way the
dual solution in string frame using 
(\ref {transfo1})-(\ref {transfo22}), (\ref {transfo3})-(\ref
{transfo4}). Starting from the latter, we can find the dual solution in
Einstein frame via the inverse conformal transformation. Of course, it is
easier to apply directly duality in Einstein frame.}.    
\vspace*{3mm} \\
The Einstein frame can be built from the string frame by the following
conformal transformation\footnote{In what follows, the tilded variables refer
to the Einstein frame whereas the ``barred'' variables are related to the dual 
solution.}
\begin{equation}
\tilde{g}_{\alpha\beta} = e^{- \phi} g_{\alpha\beta}
\label{conforme}
\end{equation}
If we introduce this transformation in the action (\ref {action1}), we 
obtain the same action, i.e. the same theory, written in the Einstein frame:
\begin{equation}
S_{eff} = \frac{1}{2 \kappa^2_D} \int{\left(
\tilde{R} - \frac{1}{12} e^{-2\phi} 
\,\tilde{H}_{\alpha\beta\delta}\,\tilde{H}^{\alpha\beta\delta}
- \frac{1}{2} \left[
\tilde{\nabla}\phi
\right]^2 + \tilde{L}_m
\right) \,\sqrt{ \mid det\,\tilde{g} \mid }\, \,d^D x}
\label{action3}
\end{equation}
where $\tilde{L}_m = e^{2 \phi}\,L_m$ and where, by convention, 
$H_{\alpha\beta\delta} = \tilde{H}_{\alpha\beta\delta}$
so that, for the dual field,
$\bar{H}_{\alpha\beta\delta} = \bar{\tilde{H}}_{\alpha\beta\delta}$ . 
\vspace*{3mm} \\
The variations of action (\ref {action3}) with respect to 
$\tilde{g}_{\alpha\beta}$, $\phi$ and $\tilde{B}_{\alpha\beta}$ give 
respectively the following field equations \cite{copel1}\footnote{Note that
in Copeland {\it et al.}'s paper \cite{copel1}, a $1/2$ factor should be 
ignored in the last term of the l.h.s. of equation (2.11).}:
\begin{equation}
\tilde{R}_{\alpha\beta} - \frac{1}{2} \tilde{g}_{\alpha\beta} \tilde{R} =
\kappa^2_D\,\tilde{T}^{(m)}_{\alpha\beta} +
\kappa^2_D\,\tilde{T}^{(\phi)}_{\alpha\beta} + 
\kappa^2_D\,\tilde{T}^{(H)}_{\alpha\beta}
\label{cheinst1}
\end{equation}
\begin{equation}
\frac{1}{6} e^{-2 \phi} \tilde{H}^{\alpha\beta\delta} 
\tilde{H}_{\alpha\beta\delta} + 
\stackrel{\sim}{\Box} \phi - \tilde{T}^{(m)}
= 0
\label{cheinst2}
\end{equation}
\begin{equation}
\tilde{\nabla}_\mu\,\left[
e^{-2 \phi} \tilde{H}^{\mu\alpha\beta}
\right] = 0
\label{cheinst3}
\end{equation}
where $\tilde{T}^{(m)}$ is $T^{(m)}_{\alpha\beta}$'s trace and
the different contributions to the energy-momentum tensor are defined by
\begin{center}
\begin{tabular}{rl}
$\kappa^2_D \tilde{T}^{(m)}_{\alpha\beta}$ & $= e^{2 \phi}\, \kappa^2_D\, 
T^{(m)}_{\alpha\beta}$ \\ 
$\kappa^2_D \tilde{T}^{(\phi)}_{\alpha\beta}$ & $= \frac{1}{2} 
\left( 
\tilde{\nabla}_\alpha \phi\,\tilde{\nabla}_\beta \phi 
- \frac{1}{2} \tilde{g}_{\alpha\beta} \left[ \tilde{\nabla} \phi \right]^2
\right)$ \\
$\kappa^2_D \tilde{T}^{(H)}_{\alpha\beta}$ & $= \frac{1}{4} e^{-2\, \phi} 
\left(
\tilde{H}_{\alpha\mu\sigma} \tilde{H}^{\hspace{3mm}\mu\sigma}_{\beta}
- \frac{1}{6} \tilde{g}_{\alpha\beta} 
\tilde{H}^{\alpha\beta\delta} \tilde{H}_{\alpha\beta\delta}
\right)$
\end{tabular}
\end{center}
In order to build up the duality transformation in the Einstein frame, we 
have to apply the conformal transformation to the duality transformation
in the string frame, i.e. to the duality transformation written in terms of 
$G$, $B$ and $T^{(m)\alpha}_{\hspace{7mm}\beta}$, and given by 
(\ref {transfo1})-(\ref {transfo22}), (\ref {transfo3})-(\ref {transfo4}). 
We obtain in this way
\begin{equation}
\begin{tabular}{rl}
$\tilde{G} \rightarrow$ & $\bar{\tilde{G}} = q\,e^{- 2 \phi}\, \tilde{P}^{-1}
$ \\ 
$\phi \rightarrow$ & $\bar{\phi}
= \phi - ln q $ \\
$\tilde{g}_{00} \rightarrow$ & $\bar{\tilde{g}}_{00} 
= q\,\tilde{g}_{00}$ \\
$\tilde{B} \rightarrow$ & $\bar{\tilde{B}} 
= - e^{- 2 \phi}\,\tilde{G}^{-1}\,\tilde{B}\,\tilde{P}^{-1}$ \\ 
$\tilde{T}^{(m)\alpha}_{\hspace{7mm}\beta} \rightarrow$ & $ 
\bar{\tilde{T}}^{(m)\alpha}_{\hspace{7mm}\beta} = \pm\,q \,
\tilde{T}^{(m)\alpha}_{\hspace{7mm}\beta}$ 
\end{tabular}
\label{transfo5}
\end{equation} 
where we have introduced the $d \times d$ matrix $\tilde{P}$ and the number 
$q$ defined by
\begin{equation}
\begin{tabular}{rl}
$\tilde{P}$ & $= \tilde{G} - e^{- 2 \phi}\,\tilde{B}\,
\tilde{G}^{-1}\,\tilde{B}$ \\ 
$q$ & $= e^{d \phi}\,\sqrt{det\,\tilde{G} \tilde{P} }$
\end{tabular}
\label{defpandq}
\end{equation}
and where we must take the ``$+$'' sign in 
$\bar{\tilde{T}}^{(m)\alpha}_{\hspace{7mm}\beta}$ if $\alpha = 0$ and the 
``$-$'' sign if $\alpha \not= 0$. 
We see that in Einstein frame, because of the presence of the conformal 
factor in the metric's definition (\ref {conforme}), $\tilde{g}_{00}$
will be modified by duality. 

\subsection{What happens when the fields depend on several coordinates ?} 
Consider now a more general case i.e. all fields ($g_{\alpha\beta}$, 
$\phi$, $B_{\alpha\beta}$ and $T^{(m)}_{\alpha\beta}$) depending on several
coordinates. Again, we shall order the coordinates so that we can say that 
all fields depend on $x^0$, ..., $x^{D-d-1}$ and do not depend on
$x^{D-d}$, ..., $x^{D-1}$. It is again necessary to suppose that the metric 
and the potential can be written as
\vspace*{2mm}
\begin{equation}
g(x^k) = \left(
\begin{tabular}{cccccc|c}
$g_{00}$ & $0$ & $0$ & ... & $0$ & $0$ & $0$ \\
$0$ & $g_{11}$ & $0$ & ... & $0$ & $0$ & $0$ \\
$0$ & $0$ & $0$ & ... & $0$ & $g_{D-d-1,D-d-1}$ & $0$\\ \\
\hline \\
& & & $0$ & & & \hspace{2mm} $G$ \\
\end{tabular}
\right)
\end{equation}
\vspace*{2mm}
\begin{equation}
B(x^k) = \left(
\begin{tabular}{l|l}
$\tilde{0}$ & $0$ \\
\hline 
$0$         & $B$
\end{tabular}
\right)
\end{equation}
\vspace*{1mm} \\ 
where $G$ and $B$ are $d \times d$ matrices whose components depend on 
$x^k$ ($k=0,...,D-d-1$) and where $\tilde{0}$ is the $ D-d \times D-d$ zero 
matrix.
\vspace*{3mm} \\
In the same way as before, we can show that, using definitions
(\ref {defphi}) and (\ref {defm}), the action (\ref {action1}) can be 
rewritten as 
$$
S_{eff} = \frac{1}{2 \kappa^2_D} \int{ e^{- \Phi}}\,\left( e^\phi\,L_m  
+ \sum_{i=0}^{D-d-1} g^{ii} \,\left\{ 
\left[ \partial_i \Phi \right]^2
+ \,\frac{1}{8}\, Tr \left[ \partial_i M \eta \partial_i M \eta \right] 
\right\}\right. \hspace{26mm} 
$$
\begin{equation}
\hspace{25mm} \left.
- \sum_{i=0}^{D-d-1} \left\{ \partial^2_i g^{ii} +              
\frac{1}{4} g^{ii}\,\left[
\partial_i g^{ii} \partial_i g_{ii} - 
\sum_{j \not= i} \partial_i g^{jj} \partial_i g_{jj} \right]
\right\} 
\right)\, d^D x
\label{action4}
\end{equation}
Again, it is easy to see that the action (\ref {action4}) is invariant under 
duality transformation given by (\ref {transfo1})-(\ref {transfo22}) and 
(\ref {transfo3})-(\ref {transfo4}), since duality does not act on the
$g^{ii}$'s components (i=0,...,D-d-1). So, duality transformation has the same
form when all fields depend on one or several coordinates, the only 
difference being the dimension of $M$ and $\eta$ matrices. 
In the Einstein frame, the duality transformation is again given by the 
transformations (\ref {transfo5}) with definitions (\ref {defpandq}). 

\subsection{A simple case: B=0.}
Since later we shall apply duality transformation to simple examples for 
which $B = 0$, it is interesting to examine the form of the duality
transformation in this case. 
\vspace*{3mm} \\
We have seen that independently of the number of coordinates the metric
tensor and the other fields depend on, the duality transformation is given 
by (\ref {transfo1})-(\ref {transfo22}) and (\ref {transfo3})-(\ref 
{transfo4}), in string frame, and by (\ref {transfo5})-(\ref {defpandq}) in 
Einstein frame. Accordingly,
we shall give the form of this transformation when $B=0$, without 
taking into account the number of coordinates all fields depend on.
\vspace*{3mm} \\
If we introduce $B=0$ in the matrix $M$ defined by (\ref {defm}), we obtain 
the following simple form for $M$:
\begin{equation}
M = \left( 
\begin{tabular}{cc}
$G^{-1}$ & $0$ \\
$0$ & $G$ 
\end{tabular}
\right)
\end{equation}
If we use transformation (\ref {mbarre}), we obtain for the matrix $\bar{M}$
\begin{equation}
\bar{M} = \left( 
\begin{tabular}{cc}
$\bar{G}^{-1}$ & $0$ \\
$0$ & $\bar{G}$ 
\end{tabular}
\right) =
\left( 
\begin{tabular}{cc}
$G$ & $0$ \\
$0$ & $G^{-1}$ 
\end{tabular}
\right)
\end{equation}
so the duality transformation in string frame takes the form
\begin{equation}
\begin{tabular}{rl}
$G \rightarrow$ & $\bar{G} = G^{-1}$ \\
$\phi \rightarrow$ & $\bar{\phi} = \phi - \,ln \left( det\,G \right)$ \\
$B = 0 \rightarrow$ & $\bar{B} = 0$ \\                 
$T^{(m)\alpha}_{\hspace{7mm}\beta} \rightarrow$ & $ 
\bar{T}^{(m)\alpha}_{\hspace{7mm}\beta} = \pm \,
\left[ det\,G \right]\, T^{(m)\alpha}_{\hspace{7mm}\beta}$ 
\end{tabular}
\label{transfo6}
\end{equation}
where we must take the ``$+$'' sign in the energy-momentum tensor if 
$\alpha = i$ and the ``$-$'' sign if $\alpha \not= i$, the i-index 
corresponding to all the coordinates the metric and the other fields depend 
on ($i = 0, ..., D-d-1$).
\vspace*{3mm} \\
In Einstein frame, the duality transformation obtained by taking $B=0$ is
\begin{equation}
\begin{tabular}{rl}
$\tilde{G} \rightarrow$ & $\bar{\tilde{G}} = q\,
e^{- 2 \phi}\,\tilde{G}^{-1}$ \\ 
$\phi \rightarrow$ & $\bar{\phi}
= \phi - ln q $ \\
$\tilde{g}_{00} \rightarrow$ & $ \bar{\tilde{g}}_{00} = q\, \tilde{g}_{00}$ \\
$\tilde{B} = 0 \rightarrow$ & $\bar{\tilde{B}} = 0$ \\
$\tilde{T}^{(m)\alpha}_{\hspace{7mm}\beta} \rightarrow$ & $ 
\bar{\tilde{T}}^{(m)\alpha}_{\hspace{7mm}\beta} = \pm\,q\,
\tilde{T}^{(m)\alpha}_{\hspace{7mm}\beta}$ 
\end{tabular}
\label{transfo7}
\end{equation} 
with $q = e^{d \phi}\,\mid\,det\,\tilde{G}\,\mid\,$ and
where we must again take the ``$+$'' sign in the energy-momentum tensor if 
$\alpha = i$ and the ``$-$'' sign, if $\alpha \not= i$ ($i = 0, ..., D-d-1$). 

\section{Examples}

\subsection{Flat FLRW metric in general relativity.}
As a first example, we shall consider the flat FLRW metric, i.e. Einstein-de
Sitter metric, that can be written as
\begin{equation}
ds^2 = - dt^2 + a(t)^2 \left[ dx^2 + dy^2 + dz^2 \right]
\end{equation}
where $t$ is the proper time and $a$ is the scale factor function of $t$. The
solution of general relativity field equations in presence of perfect fluid  
with state equation $p = \rho / 3$ (radiation) is given
by\footnote{In all the following examples, $D = 4$ so that $\kappa^2_D =
\kappa^2_4$. But, for simplicity, we shall note $\kappa^2$ instead of
$\kappa^2_4$.} 
\begin{equation}
\begin{tabular}{rl}
$a(t)$              & $ = t^{1/2}$ \\ 
$\kappa^2\,\rho(t)$ & $ = 3/4\, t^{-2}$ \\
$\kappa^2\,p(t)$    & $ = 1/4\,t^{-2}$
\end{tabular}
\label{sol4}
\end{equation}
It is important to note that one recovers the general relativity limit from 
(\ref {eqncordes})-(\ref {eqncordes3}) only when $H = 0$, $\phi = 0$ and 
$T^{(m)} = 0$, where $T^{(m)}$ is the energy-momentum tensor's trace, so that
in this case only can a solution of Einstein field equations with a perfect
fluid be at the same time a solution of string field equations. 
So (\ref {sol4}) being a general relativity solution with $T^{(m)} = 0$, is 
also solution of string theory field equations in string as well as
in Einstein frames. This metric is homogeneous and written in terms of proper time
($g_{00}=-1$) so our duality transformation in string frame given by 
(\ref {transfo1})-(\ref {transfo22}) and (\ref {transfo3})-(\ref {transfo4}) 
reduces to the one given in \cite{vene1}. The matrix $G$ can be written as
\begin{equation}
G = \left(
\begin{tabular}{ccc}
$a^2(t)$ & $0$ & $0$ \\
$0$ & $a^2(t)$ & $0$ \\
$0$ & $0$ & $a^2(t)$ 
\end{tabular}
\right)
\label{redefg}
\end{equation}
Introducing this matrix and (\ref {sol4}) in (\ref {transfo6}), we find 
the corresponding dual solution in string frame:
\begin{equation}
\begin{tabular}{rl}
$\bar{a}(t)$              & $ = t^{-1/2}$ \\ 
$\bar{\phi}(t)$           & $ = - 3\,ln(t) $ \\
$\kappa^2\,\bar{\rho}(t)$ & $ = 3/4\,t$ \\
$\kappa^2\,\bar{p}(t)$    & $ = - 1/4\,t$
\end{tabular}
\label{sol5}
\end{equation}
where the dual metric is 
\begin{equation}
ds^2 = - dt^2 + \bar{a}(t)^2 \left[ dx^2 + dy^2 + dz^2 \right]   
\end{equation}
To find the dual solution in Einstein frame, one only has to apply 
transformation
(\ref {transfo7}) to (\ref {sol4}) and (\ref {redefg}) (as we are in general
relativity, we have $\tilde{G} = G$). So the dual solution in Einstein frame
is
\begin{equation}
\begin{tabular}{rl}
$\bar{\tilde{a}}(t)$      & $= t$ \\
$\bar{\tilde{g_{00}}}(t)$ & $= - t^3$ \\
$\bar{\phi}(t)$           & $= - 3 \,ln t$ \\
$\kappa^2\,\bar{\rho}(t)$ & $= 3/4\,t$ \\
$\kappa^2\,\bar{p}(t)$    & $= - 1/4\,t$
\end{tabular}
\label{sol6}
\end{equation}
The dual metric found in this way is 
\begin{equation}
ds^2 = - t^3 dt^2 + \bar{\tilde{a}}^2(t) 
\left[ dx^2 + dy^2 + dz^2 \right]
\end{equation}
We see that, in the Einstein frame, starting from a solution written in terms
of proper time, the dual solution is no more expressed in terms of proper
time.  If we want to write it in terms of proper time, it is necessary to change the time-like
coordinate from $t$ to $\tilde{t}$ so that $\tilde{t}$ defined by
\begin{equation}
\tilde{t} = \int{t^{\,3/2}\,dt} \propto t^{\,5/2}
\end{equation}
is the new proper time in Einstein frame.
We can finally write the dual solution in terms of proper time:
\begin{itemize}
\item the dual metric
\begin{equation}
ds^2 = - d\tilde{t}^2 + \bar{\tilde{a}}^2 
\left[ dx^2 + dy^2 + dz^2 \right]
\end{equation} 
with $\bar{\tilde{a}}(\tilde{t}) = \tilde{t}^{\,2/5}$
\item the dual fields
\begin{equation}
\begin{tabular}{rl}
$\bar{\phi}(\tilde{t})$           & $ = - 6/5 \,ln \tilde{t}$ \\
$\kappa^2\,\bar{\rho}(\tilde{t})$ & $ = 3/4\, \tilde{t}^{\,2/5}$ \\
$\kappa^2\,\bar{p}(\tilde{t})$    & $ = - 1/4\,\tilde{t}^{\,2/5}$
\end{tabular}
\end{equation}
\end{itemize}
Note that duality transformation applied to a solution written in the proper
time of Einstein frame always implies a proper time redefinition.
 
\subsection{Bianchi I metric in general relativity.}
As a second example, we shall consider the Bianchi I cosmological
solution of general relativity with perfect fluid \cite{moi}. 
\vspace*{3mm} \\
The Bianchi I metric takes the following form  
\begin{equation}
ds^2 = - (a b c)^2 d\eta^2 + a^2 dx^2 + b^2 dy^2 + c^2 dz^2
\label{metricbch1}
\end{equation}
where $a$, $b$ and $c$ are the scale factors functions of $\eta$. We do not 
use in (\ref {metricbch1}) proper time but  logarithmic time
denoted by $\eta$ and related to proper time $t$ by
\begin{equation}
dt = a\,b\,c\,d\eta
\end{equation}
This metric describes a four-dimensional homogeneous cosmological
background so $x^0$ is the time-like coordinate and the energy-momentum 
tensor can be written as $T^{(m)\alpha}_{\hspace{7mm}\beta} = 
diag(-\rho, p_1, p_2, p_3)$\footnote{In all the 
following examples, we shall consider a diagonal energy-momentum tensor.}
where $\rho(\eta)$ is the energy-density and $p_1$, $p_2$ and $p_3$ are the 
three pressure's components. 
We shall take for state equation of the perfect fluid $p = \gamma \rho$ where
we shall restrict ourselves to $0 \leq \gamma < 1$.
\vspace*{3mm} \\ 
It is important to remind that in section 2 duality transformation has been 
es\-ta\-bli\-shed without making any hypothesis about the time-like coordinate
to be used. Indeed duality transformation can be applied for any
function $g_{00}(x^0)$: it is not necessary that $g_{00} = -1$ as in the case
of proper time.
\vspace*{3mm} \\ 
The solution of general relativity field equations with a perfect fluid 
characterized by an isotropic pressure ($p = p_1 = p_2 = p_3$), i.e.
\begin{equation}
R_{\alpha\beta} - \frac{1}{2}\,g_{\alpha\beta}\,R = \kappa^2
T^{(m)}_{\alpha\beta} 
\label{eqnrg}
\end{equation}
is displayed below\footnote{This solution
is known in closed form in terms of proper time for $\gamma = 0$ only \cite{jacobs}.}
\cite{moi}
\begin{equation}
a(\eta) = e^{(\beta_0 + \delta_0)\,\eta / 3}\, sinh^{-2/3/(1-\gamma)}\left[
\frac{1}{2}\, \sqrt{\epsilon}\,(1 - \gamma)\,\eta \right]
\end{equation}
\begin{equation}
b(\eta) = e^{(\delta_0 - 2 \beta_0)\,\eta / 3}\, sinh^{-2/3/(1-\gamma)}\left[
\frac{1}{2}\, \sqrt{\epsilon}\,(1 - \gamma)\,\eta \right]
\end{equation}
\begin{equation}
c(\eta) = e^{(\beta_0 - 2 \delta_0)\,\eta / 3}\, sinh^{-2/3/(1-\gamma)}\left[
\frac{1}{2}\, \sqrt{\epsilon}\,(1 - \gamma)\,\eta \right]
\end{equation}
\begin{equation}
\kappa^2 \rho(\eta) = \frac{\epsilon}{3} \,sinh^{-2\omega} \left[
\frac{1}{2}\,\sqrt{\epsilon}\,(1-\gamma)\,\eta \right] \hspace{20mm}
\end{equation}
\begin{equation}
\kappa^2 p(\eta) = \frac{\gamma \epsilon}{3} \,sinh^{-2\omega} \left[
\frac{1}{2}\,\sqrt{\epsilon}\,(1-\gamma)\,\eta \right] \hspace{20mm}
\end{equation}
where $\omega$, $\epsilon$, $\beta_0$ and $\delta_0$ are constants related by 
\begin{equation}
\omega = \frac{\gamma + 1}{\gamma - 1} \hspace{17mm}
\end{equation}
and
\begin{equation}
\epsilon = \beta_0^2 + \delta_0^2 - \beta_0 \delta_0
\end{equation}
The duality transformation can be applied to a string theory solution only. 
The above solution is also a string theory solution only when the
energy-momentum tensor's trace is null, so when $p=\rho/3$. Indeed it can 
easily be checked that the solution obtained with $\gamma = 0$ (dust 
universe) is not a solution of equations (\ref {eqncordes})-(\ref 
{eqncordes3}) contrary to the solution with $\gamma = 1/3$. 
As a solution in explicit form for $\gamma = 1/3$ cannot be found in terms of
proper time but only in terms of logarithmic time, as far as we know, we have
to resort to our duality transformation given in the preceding section.
\vspace*{3mm} \\
For $\gamma = 1/3$, the above 
solution takes the following form
\begin{equation}
a(\eta) = e^{(\beta_0 + \delta_0)\eta / 3}\, sinh^{-1}\left[
\frac{1}{3}\, \sqrt{\epsilon}\,\eta \right]
\end{equation}
\begin{equation}
b(\eta) = e^{(\delta_0 - 2 \beta_0)\eta / 3}\, sinh^{-1}\left[
\frac{1}{3}\, \sqrt{\epsilon}\,\eta \right]
\end{equation}
\begin{equation}
c(\eta) = e^{(\beta_0 - 2 \delta_0)\eta / 3}\, sinh^{-1}\left[
\frac{1}{3}\, \sqrt{\epsilon}\,\eta \right]
\end{equation}
\begin{equation}
\kappa^2 \rho(\eta) = \frac{\epsilon}{3} \,sinh^4 \left[
\frac{1}{3}\,\sqrt{\epsilon}\,\eta \right] \hspace*{20mm}
\end{equation}
\begin{equation}
\kappa^2 p(\eta) = \frac{\epsilon}{9} \,sinh^4 \left[
\frac{1}{3}\,\sqrt{\epsilon}\,\eta \right] \hspace*{20mm}
\end{equation}
with 
\begin{equation}
\epsilon = \beta_0^2 + \delta_0^2 - \beta_0 \delta_0
\end{equation}
This solution is a string theory solution in string as well as in Einstein 
frames. 
As the metric is four-dimensional and homogeneous, the matrix $G$ defined by 
(\ref {defgandb}) can be written as
\begin{equation}
G = \left( 
\begin{tabular}{ccc}
$a(\eta)^2$ & $0$ & $0$ \\
$0$ & $b(\eta)^2$ & $0$ \\
$0$ & $0$ & $c(\eta)^2$
\end{tabular}
\right)
\end{equation}
and using this matrix $G$, the relations (\ref {transfo6}) enable one to find 
the dual solution in string frame whereas the relations (\ref {transfo7}) 
with $\tilde{G} = G$ give the dual solution in Einstein frame, i.e. 
respectively
\begin{equation}
\bar{a}(\eta) = e^{- (\beta_0 + \delta_0)\eta / 3}\, sinh \left[
\frac{1}{3}\, \sqrt{\epsilon}\,\eta \right] \hspace*{8mm}
\end{equation}
\begin{equation}
\bar{b}(\eta) = e^{(2 \beta_0 - \delta_0)\eta / 3}\, sinh \left[
\frac{1}{3}\, \sqrt{\epsilon}\,\eta \right] \hspace*{8mm}
\end{equation}
\begin{equation}
\bar{c}(\eta) = e^{(2 \delta - \beta_0)\eta / 3}\, sinh \left[
\frac{1}{3}\, \sqrt{\epsilon}\,\eta \right] \hspace*{8mm}
\end{equation}
\begin{equation}
\bar{\tilde{a}}(\eta) = e^{- (\beta_0 + \delta_0)\eta / 3}\, sinh^{-2}\left[
\frac{1}{3}\, \sqrt{\epsilon}\,\eta \right]
\end{equation}
\begin{equation}
\bar{\tilde{b}}(\eta) = e^{- (\delta_0 - 2 \beta_0)\eta / 3}\, sinh^{-2}\left[
\frac{1}{3}\, \sqrt{\epsilon}\,\eta \right]
\end{equation}
\begin{equation}
\bar{\tilde{c}}(\eta) = e^{- (\beta_0 - 2 \delta_0)\eta / 3}\, sinh^{-2}\left[
\frac{1}{3}\, \sqrt{\epsilon}\,\eta \right]
\end{equation}
\begin{equation}
\bar{\phi}(\eta) = 6\, ln \left( sinh \left[
\frac{1}{3}\, \sqrt{\epsilon}\,\eta \right] \right) \hspace*{12mm}
\end{equation}
\begin{equation}
\kappa^2 \bar{\rho}(\eta) = \frac{\epsilon}{3} \,sinh^{-2} \left[
\frac{1}{3}\,\sqrt{\epsilon}\,\eta \right] \hspace*{12mm}
\end{equation}
\begin{equation}
\kappa^2 \bar{p}(\eta) = - \frac{\epsilon}{9} \,sinh^{-2} \left[
\frac{1}{3}\,\sqrt{\epsilon}\,\eta \right] \hspace*{12mm}
\end{equation}
with
\begin{equation}
\epsilon = \beta_0^2 + \delta_0^2 - \beta_0 \delta_0
\end{equation}
where the tilded scale factors are related to Einstein frame whereas the 
untilded ones refer to string frame.
\vspace*{3mm} \\
As is manifest in this example, fluid's pressure changes sign after
duality but not its density.

\subsection{Bianchi I metric with a scalar field.}
We shall consider again the metric (\ref {metricbch1}) but we shall apply 
duality transformation to the string theory solution with $\phi \not= 0$. The
solution with a perfect fluid with state equation $p = \gamma \rho$ 
(with $0 \leq \gamma < 1$) of equations (\ref{eqncordes})-(\ref {eqncordes2})
is given by \cite{moi}
\begin{itemize}
\item The scale factors in string frame
\begin{equation}
\begin{tabular}{|rl}
$a(\eta)$ & $= e^{(D- A B)\,\eta/2} 
sinh^{k+\omega/2} \left[ \frac{1}{2}\,\sqrt{\epsilon}\,(1-\gamma)\,\eta
\right]$ \\
$b(\eta)$ & $= e^{(D- A B-2\beta_0)\,\eta/2} 
sinh^{k+\omega/2} \left[ \frac{1}{2}\,\sqrt{\epsilon}\,(1-\gamma)\,\eta
\right]$ \\
$c(\eta)$ & $= e^{(D- A B -2\delta_0)\,\eta/2} 
sinh^{k+\omega/2} \left[ \frac{1}{2}\,\sqrt{\epsilon}\,(1-\gamma)\,\eta
\right]$
\end{tabular}
\end{equation}
\item The scale factors in Einstein frame 
\begin{equation}
\begin{tabular}{|rl}
$\tilde{a}(\eta)$ & $= e^{-A B \eta / 2} 
sinh^k \left[ \frac{1}{2}\,\sqrt{\epsilon}\,(1-\gamma)\,\eta\right]$ 
\hspace*{5mm}\\
$\tilde{b}(\eta)$ & $= e^{-(A B / 2 + \beta_0)\, \eta} 
sinh^k \left[ \frac{1}{2}\,\sqrt{\epsilon}\,(1-\gamma)\,\eta\right]$ \\
$\tilde{c}(\eta)$ & $= e^{-(A B / 2 + \delta_0) \,\eta} 
sinh^k \left[ \frac{1}{2}\,\sqrt{\epsilon}\,(1-\gamma)\,\eta\right]$
\end{tabular}
\end{equation}
\item The other fields
\begin{equation}
\begin{tabular}{|rl}
$\phi(\eta)$ & $= D \eta + \omega\, ln \left(
sinh\left[ \frac{1}{2}\,\sqrt{\epsilon}\,(1-\gamma)\,\eta\right] \right)$ \\
$\kappa^2 \rho(\eta)$ & $= \epsilon\,A\,e^{E\eta} sinh^l\left[
\frac{1}{2}\,\sqrt{\epsilon}\,(1-\gamma)\,\eta \right]$ \\
$\kappa^2 p(\eta)$ & $= \epsilon \gamma\,A\,e^{E\eta} sinh^l\left[
\frac{1}{2}\,\sqrt{\epsilon}\,(1-\gamma)\,\eta \right]$
\end{tabular}
\end{equation}
\end{itemize}
with 
$$
A = \left[ 3 - \left( 
\frac{3\gamma-1}{1-\gamma}\right)^2\right]^{-1}
$$ 
$$
B = - 2 \left( \beta_0 + \delta_0 + 
\frac{\phi_0}{2}\,\frac{3\gamma-1}{1-\gamma}\right) 
$$
$$
C = \beta_0 \delta_0 - \frac{1}{4} \phi_0^2 
$$
$$
D = \phi_0 - \frac{B}{A}\,\left( \frac{3\gamma-1}{1-\gamma} \right) 
$$
$$ 
E = 2 \left( \frac{B(1+3\gamma)}{2A(1-\gamma)} + \beta_0 + \delta_0
- \phi_0 \right)
$$
$$ 
k = - \frac{2}{A(1-\gamma)}
$$
$$ 
\omega = \frac{-4}{A}\left(\frac{3\gamma-1}{(1-\gamma)^2} \right)
$$
$$
\epsilon = \frac{B^2}{4} - AC
$$
$$
l = - 2 + \frac{4}{A}\, \frac{3\gamma+1}{(1-\gamma)^2}
$$
Contrary to the previous example of Bianchi I metric in general relativity,
we can apply here the duality transformation to the above solution for any  
$0 \leq \gamma < 1$: this is so because we are in string theory instead of in
general relativity. 
\vspace*{3mm} \\
We shall give e.g. the explicit solution for dust fluid ($\gamma = 0$):
\begin{equation}
\begin{tabular}{l}
$a(\eta) = e^{\phi_0 \eta / 2}$ \\
$b(\eta) = e^{(\phi_0 - 2 \beta_0)\, \eta / 2}$ \\
$c(\eta) = e^{(\phi_0 - 2 \delta_0)\, \eta / 2}$ \\
$\tilde{a}(\eta) = e^{(\beta_0 + \delta_0 - \phi_0/2)\, \eta / 2}\,
sinh^{-1}\left[ \frac{1}{2}\,\sqrt{\epsilon}\,\eta \right]$ \\
$\tilde{b}(\eta) = e^{(- \beta_0 + \delta_0 - \phi_0/2)\, \eta / 2}\,
sinh^{-1}\left[ \frac{1}{2}\,\sqrt{\epsilon}\,\eta \right]$ \\
$\tilde{c}(\eta) = e^{(\beta_0 - \delta_0 - \phi_0/2)\, \eta / 2}\,
sinh^{-1}\left[ \frac{1}{2}\,\sqrt{\epsilon}\,\eta \right]$ \\
$\phi(\eta) = (3 \phi_0/2 - \beta_0 - \delta_0)\,\eta + 2\,ln \left(
sinh \left[ \frac{1}{2}\,\sqrt{\epsilon}\,\eta \right] \right)$ \\
$\kappa^2 \rho(\eta) = \frac{\epsilon}{2}\,
e^{(\beta_0 + \delta_0 - 3 \phi_0 /2)\,\eta}$ \\
$\kappa^2 p(\eta) = 0$ \\
\end{tabular}
\end{equation}
with as constraint
\begin{equation}
\epsilon = \beta_0^2 + \delta_0^2 - \phi_0 \left( \beta_0 + \delta_0 \right) +
\frac{3}{4} \phi_0^2
\label{constraint2}
\end{equation}
The dual solution can be found by applying to the above solution transformation
(\ref {transfo6}) in string frame and transformation (\ref {transfo7}) in
Einstein frame. We obtain in this way
\begin{equation}
\begin{tabular}{l}
$\bar{a}(\eta) = e^{- \phi_0 \eta / 2}$ \\
$\bar{b}(\eta) = e^{(\beta_0 - \phi_0/2)\, \eta}$ \\
$\bar{c}(\eta) = e^{(\delta_0 - \phi_0/2)\, \eta}$ \\
$\bar{\tilde{a}}(\eta) = e^{- (\beta_0 + \delta_0 - \phi_0/2)\, \eta / 2}\,
sinh^{-1}\left[ \frac{1}{2}\,\sqrt{\epsilon}\,\eta \right]$ \\
$\bar{\tilde{b}}(\eta) = e^{(\beta_0 - \delta_0 + \phi_0/2)\, \eta / 2}\,
sinh^{-1}\left[ \frac{1}{2}\,\sqrt{\epsilon}\,\eta \right]$ \\
$\bar{\tilde{c}}(\eta) = e^{(- \beta_0 + \delta_0 + \phi_0/2)\, \eta / 2}\,
sinh^{-1}\left[ \frac{1}{2}\,\sqrt{\epsilon}\,\eta \right]$ \\
$\bar{\phi}(\eta) = (- 3 \phi_0/2 + \beta_0 + \delta_0)\,\eta + 2 ln \left(
sinh \left[ \frac{1}{2}\,\sqrt{\epsilon}\,\eta \right] \right)$ \\
$\kappa^2 \bar{\rho}(\eta) = \frac{\epsilon}{2}\,
e^{- (\beta_0 + \delta_0 - 3 \phi_0 /2)\,\eta} $ \\
$\kappa^2 \bar{p}(\eta) = 0$ \\
\end{tabular}
\end{equation}
with again (\ref {constraint2}) as constraint.

\subsection{Texeira et al.'s inhomogeneous metric.}
For the following example, we have chosen to perform duality on an 
inhomogeneous metric that has the following form \cite{texeira}
\begin{equation}
ds^2 = - e^{2 \nu(z)}\,dt^2 + z^2 \left[ dx^2 + dy^2 \right] 
+ \frac{z}{F(z)}\,dz^2
\label{texeira}
\end{equation}
The exact solution of the corresponding general relativistic field equations
for this space-time filled with a perfect fluid with state equation 
$p = \rho / 3$ is well known \cite{texeira}:
\begin{equation}
\begin{tabular}{ll}
$\nu(z)$  & $= - \frac{1}{2} ln(z) - \frac{1}{2} ln (6 - z^5)$ \\ \\
$F(z)$    & $= \frac{1}{5}\, (6 - z^5)^3$ \\ \\
$\kappa^2 p_1(z)$ & $= \kappa^2 p_2(z) = \kappa^2 p_3(z) = z^2 (6 - z^5)^2$
\\ \\
$\kappa^2 \rho(z)$ & $= 3 \,z^2 (6 - z^5)^2$ 
\end{tabular}
\label{sol}
\end{equation}
\underline{Case 1.} \vspace*{3mm} \\
As the metric is four-dimensional and depends on z-coordinate only, the 
matrix $G$ defined by (\ref {defgandb}) can be written as 
\begin{equation}
G = \left(
\begin{tabular}{ccc}
$- e^{2 \nu(z)}$ & $0$   & $0$ \\
$0$           & $z^2$ & $0$ \\
$0$           & $0$   & $z^2$
\end{tabular}
\right)
\label{defg2}
\end{equation}
Using the transformation given by (\ref {transfo6}), we find the dual solution
in string frame, under the following form
\begin{equation}
ds^2 = - e^{- 2 \nu(z)}\,dt^2 + \frac{1}{z^2} \left[ dx^2 + dy^2 \right] + 
\frac{z}{F(z)}\, dz^2
\end{equation}
The corresponding dual scalar fields can be written as
\begin{equation}
\begin{tabular}{rl}
$\bar{\phi}(z)$ & $= - 3\,ln(z) + ln (6 - z^5)$ \\
$\kappa^2\,\bar{p}_1(z)$  & $= - z^5 (6 - z^5)$ \\
$\kappa^2\,\bar{p}_2(z)$  & $= - z^5 (6 - z^5)$ \\
$\kappa^2\,\bar{p}_3(z)$  & $= z^5 (6 - z^5)$ \\
$\kappa^2\,\bar{\rho}(z)$ & $= - 3 z^5 (6 - z^5)$
\end{tabular}
\end{equation}
Some remarks have to be done here. First we see that starting from a solution
with isotropic pressure, we find a dual solution with anisotropic pressure:
two components are negative and one remains positive (the one corresponding to
the z-coordinate), but in absolute value, the pressure remains the same in
the three directions. Secondly, we note that after performing the duality
transformation, the energy
density becomes negative and so we are led to question the physical validity 
of this dual solution.
\vspace*{3mm} \\
\underline{Case 2.}
\vspace*{3mm} \\
It is interesting to note that as $B = 0$, we are free to choose $G$'s 
dimension. Indeed, (\ref {defg2}) is the matrix $G$ with maximal dimension 
but as the potential $B$ does not appear in the duality transformation, we 
may choose a matrix $G$ with a smaller dimension. For example, we can
write the metric (\ref {texeira}) as
\begin{equation}
g_{\alpha\beta} = \left(
\begin{tabular}{cc|c}
$z/F(z)$ & $0$ & $0$ \\
$0$ & $-e^{2 \nu}$ & $0$ \\
\hline
$0$ & $0$ & $G(z)$
\end{tabular}
\right)
\end{equation}
with
\begin{equation}
G = \left(
\begin{tabular}{cc}
$z^2$ & $0$ \\
$0$   & $z^2$
\end{tabular}
\right)
\label{matrixg}
\end{equation}
and we can perform the duality transformation on the solution (\ref {sol}) 
with (\ref {matrixg}) as matrix $G$. Again duality transformation is given by 
(\ref {transfo6}) in string frame and the dual solution obtained is
\begin{itemize}
\item for the metric: 
\begin{equation}
ds^2 = - e^{2 \nu(z)}\,dt^2 + \frac{1}{z^2} \left[ dx^2 + dy^2 \right] + 
\frac{z}{F(z)}\, dz^2
\end{equation}
\item for the scalar fields: 
\begin{equation}
\begin{tabular}{rl}
$\bar{\phi}(z)$ & $= - 4\,ln(z)$ \\
$\kappa^2\,\bar{\rho}(z)$ & $= 3\, z^6 (6 - z^5)^2$ \\
$\kappa^2\,\bar{p}_1(z)$  & $= - z^6 (6 - z^5)^2$ \\
$\kappa^2\,\bar{p}_2(z)$  & $= - z^6 (6 - z^5)^2$ \\
$\kappa^2\,\bar{p}_3(z)$  & $= z^6 (6 - z^5)^2$ 
\end{tabular}
\end{equation}
\end{itemize}
As we have removed $g_{00}$ from the matrix $G$ given by (\ref {matrixg}) 
with respect to its form (\ref {defg2}), we can see that the energy density 
remains positive. In fact, we can say in general that if $g_{ii}$ is present
in $G$, then the corresponding energy-momentum tensor's component
$T^{(m)}_{ii}$ will change sign after duality while if $g_{ii}$ is not 
included in $G$, 
then the corresponding $T^{(m)}_{ii}$ will keep the same sign after duality. 
Indeed, in the first case ($G$ given by (\ref {defg2})),
$g_{00}$, $g_{11}$ and $g_{22}$ are present in $G$, so $\rho$, $p_1$ and
$p_2$ become negative after duality whereas $p_3$ remains positive while in
the second case ($G$ given by (\ref {matrixg})), only $g_{11}$
and $g_{22}$ are present in $G$ and so only $p_1$ and $p_2$ become negative.
\vspace*{3mm} \\
In the same way, we can perform duality with a $1 \times 1$ matrix for $G$,
for example: 
\begin{equation}
G = \left( e^{2 \nu} \right)
\end{equation}
or
\begin{equation}
G = \left( z^2 \right)                          
\end{equation}
The only way to keep an isotropic pressure after duality is to include all 
spatial metric components in $G$, i.e. in the case of a homogeneous
metric\footnote{
In fact, there is another mathematically possible case for which the pressure 
remains isotropic after duality, i.e. when none of the spatial metric 
components is in $G$, so that $g_{00}$ is the only element of $G$.
However, the energy density $\rho$ becomes negative which casts some doubt on
the physical validity of the corresponding dual solution.}.
                     
\subsection{Inhomogeneous Senovilla's metric.}
Now we shall apply duality transformation to a metric depending on several
coordinates, i. e. inhomogeneous Senovilla's metric given by \cite{senovilla}
\begin{equation}
ds^2 = e^{2 f} (- dt^2 + dx^2) + h (q dy^2 + q^{-1} dz^2)
\label{metriquesenovilla}
\end{equation}
where $f$, $h$ et $q$ are functions of $t$ and $x$. The solution found in
general relativity in presence of a perfect fluid with state equation 
$p = \rho / 3$ has been given in \cite{senovilla}:
\begin{equation}
\begin{tabular}{rl}
$e^{f(x,t)}$ & $= cosh^2(a t)\,cosh(3 a x)$ \\
$h(x,t)$   & $= cosh(a t)\,sinh(3 a x)\,cosh^{-2/3}(3 a x)$ \\
$q(x,t)$   & $= cosh^3(a t)\,sinh(3 a x)$ 
\end{tabular}
\label{solution}
\end{equation}
\begin{equation}
\begin{tabular}{rl}
$\kappa^2\,\rho(x,t)$ 
& $= 15 a^2\,\left[cosh(a t)\,cosh(3 a x) \right]^{-4}$\\ \\
$\kappa^2\,p(x,t)$   
& $= 5 a^2\,\left[cosh(a t)\,cosh(3 a x) \right]^{-4}$
\end{tabular}
\end{equation}
where $a$ is an integration constant. \vspace*{3mm} \\
As the metric depends on two coordinates, $t$ and $x$, the matrix $G$ can be
written as 
\begin{equation}
G = \left(
\begin{tabular}{cc}
$h q$ & $0$ \\
$0$   & $h q^{-1}$
\end{tabular}
\right)
\label{matriceg}
\end{equation}
Using relations given in (\ref {transfo6}), which constitute the duality 
transformation in string frame, we find the dual metric: 
\begin{equation}
ds^2 = e^{2 f} (- dt^2 + dx^2) + h^{-1} (q^{-1} dy^2 + q dz^2)
\end{equation}
with $f(x,t)$, $h(x,t)$ and $q(x,t)$ given by (\ref {solution}),
and the dual scalar fields:
\begin{equation}
\begin{tabular}{rl}
$\bar{\phi}(x,t) =$ & $ - 2 \,\left(
ln\left[cosh(a t)\right] + ln\left[sinh(3 a x)\right] - \frac{2}{3} 
ln\left[cosh(3 a x)\right]
\right)$ \\ \\
$\kappa^2\,\bar{\rho}(x,t) = $ & $
15 a^2 h^2(x,t) \left[cosh(a t) cosh(3 a x) \right]^{-4}$ \\ \\ 
$\kappa^2\,\bar{p}_1(x,t) = $ & $
5 a^2 h^2(x,t) \left[cosh(a t) cosh(3 a x) \right]^{-4}$ \\ \\
$\kappa^2\,\bar{p}_2(x,t) = $ & $ - 
5 a^2 h^2(x,t) \left[cosh(a t) cosh(3 a x) \right]^{-4}$ \\ \\
$\kappa^2\,\bar{p}_3(x,t) = $ & $ - 
5 a^2 h^2(x,t) \left[cosh(a t) cosh(3 a x) \right]^{-4}$ \\ \\
\end{tabular}
\label{scalar}
\end{equation}
where the function $h(x,t)$ is given by (\ref {solution}). \vspace*{3mm} \\
We note that since only $g_{22}$ and $g_{33}$ are present in matrix $G$, only
$p_2$ and $p_3$ change sign after duality while $\rho$ and $p_1$ remain
positive. 
If we apply transformation (\ref {transfo7}) to
(\ref {metriquesenovilla}), we find the dual metric in Einstein frame: 
\begin{equation}
ds^2 = e^{2 f} h^2\,( - dt^2 + dx^2 ) + h ( q^{-1} dy^2 + q dz^2 )
\label{metricdual}
\end{equation}
with $f(x,t)$, $h(x,t)$ and $q(x,t)$ given by (\ref {solution}), the dual 
scalar fields (\ref {scalar}) being the same in the two frames.

\subsection{Mars' inhomogeneous non-diagonal metric.}
We shall now consider the inhomogeneous non-diagonal metric 
\cite{mars}\footnote{In this paper, M.Mars does not introduce directly the 
dilatonic field: he works with a stiff fluid. But we know that a massless 
scalar field is equivalent to a stiff fluid: we can pass from the scalar 
field expression to the fluid expression by the relation: 
$p = \rho = \dot{\phi}^2 /4$ \cite{ince}.} given by
$$
ds^2 = e^{h\,t} \left( 
e^{s\,r^2}\,cosh(2\,k\,t)\,\left[- dt^2 + dr^2 \right] + \hspace{30mm}
\right.
$$ 
\begin{equation}
\left.
r^2\, 
cosh(2\,k\,t)\,d\varphi^2 + \frac{1}{cosh(2\,k\,t)}\,
(dz + k\,r^2 d\varphi)^2 \right)
\end{equation}
where $h$, $k$ and $s$ are constants. \\
This metric is solution of string field equations 
(\ref {eqncordes})-(\ref {eqncordes3}) in the absence of perfect fluid with a
dilatonic field given by
\begin{equation}
\phi = h\,t
\end{equation}
with 
\begin{equation}
h^2 + 4 k^2 - 4 s = 0
\label{contrainte}
\end{equation}
as constraint. 
As the metric depends on the two coordinates $t$ and $r$, the $G$ matrix can be
written as 
\begin{equation}
G = \frac{e^{h\,t}}{cosh(2\,k\,t)} \,\left(
\begin{tabular}{cc}
$r^2 cosh^2(2\,k\,t) + k^2 r^4$ & $k r^2$ \\
$k r^2$ & $1$
\end{tabular}
\right)
\end{equation}
In this case, $B$, $p$ and $\rho$ are absent and the duality transformation 
(\ref {transfo6}) takes the following form 
\begin{equation}
\begin{tabular}{rl}
$G$    & $\rightarrow \bar{G} = G^{-1}$ \\
$\phi$ & $\rightarrow \bar{\phi} = \phi - ln (det\,G)$
\end{tabular}
\label{transfo8}
\end{equation}
Using these relations, we can write for the matrix $\bar{G}$
\begin{equation}
\bar{G} = G^{-1} = \frac{e^{- h\,t}}{cosh(2\,k\,t)} \,\left(
\begin{tabular}{cc}
$r^{-2}$ & $- k$ \\ \\
$- k $   & $cosh^2(2\,k\,t) + k^2 r^2$ 
\end{tabular}
\right)
\end{equation}
so that the dual metric becomes
\begin{equation}
\begin{tabular}{rl}
$ds^2 =$ & $e^{h t + s r^2}\,cosh(2 k t) \left[ -dt^2 + dr^2 \right]$ \\ \\
         & $+ \left[ \left(r^{-1}\,d\varphi - k r dz \right)^2 
         + cosh^2(2 k t)\,dz^2 \right]\,\displaystyle{\frac{e^{-h t}}{cosh(2kt)}}$
\end{tabular}
\end{equation}
Finally the dilatonic field can be written as 
\begin{equation}
\bar{\phi} = - h\,t - 2\, ln r
\end{equation}
the constraint (\ref {contrainte}) remaining still valid.

\subsection{Schwarzschild's metric in general relativity.} 
The last example we shall consider is Schwarzschild's metric in general 
relativity\footnote{As it is a vacuum solution, it is 
also a string theory solution and we can so perform on it the duality 
transformation.}: 
\begin{equation}
ds^2 = - e^{\lambda(r)} dt^2 + e^{\nu(r)} dr^2 + r^2 d\theta^2 
+ r^2 sin^2 \theta d\varphi^2
\label{schwarzschild}              
\end{equation}
with 
\begin{equation}
e^{\lambda(r)} = e^{-\nu(r)} = 1 - \frac{2 G M}{r} 
\end{equation}
This solution is given for $\phi = p = \rho = 0$ and is different from the
preceding examples since the metric is now written in spherical coordinates.
As this metric depends on two coordinates ($r$, $\theta$), we could
be tempted to choose the following matrix $G$
\begin{equation}
G = \left(
\begin{tabular}{cc}
$e^\lambda$ & $0$ \\
$0$         & $r^2 sin^2\theta$
\end{tabular}
\right)
\end{equation}
If we apply the duality transformation (\ref {transfo8}) to the above matrix
$G$, we realize that the dual solution obtained in this way is not a solution
of field equations (\ref {eqncordes})-(\ref {eqncordes3}). The fundamental
reason for this is due to the fact that the spatial metric is written in 
spherical coordinates. If we rewrite the metric in Cartesian coordinates, we
see that, in fact, it depends on the three space coordinates. Thus the matrix
$G$ has in fact to be chosen as the one-dimensional matrix 
\begin{equation}
G = \left( e^\lambda \right)
\end{equation}
Using the duality transformation (\ref {transfo8}), we find the corresponding
dual solution 
\begin{equation}
\begin{tabular}{rl}
$\bar{G}$    & $= \left( e^{-\lambda} \right)$ \\
$\bar{\phi}$ & $= - \lambda$
\end{tabular}
\end{equation}
so that the genuine dual metric, back in spherical coordinates, can be written as
\begin{equation}
ds^2 = e^{- \lambda} \left[ - dt^2 + dr^2 \right] 
+ r^2 \left[ d\theta^2 + sin^2\theta d\varphi^2 \right]
\end{equation}
Note that, in this way, we do not modify the angular components of the 
metric, so we keep the spherical symmetry of the initial Schwarzschild
solution.

\section{Some future perspectives...}
The main use of the dual transformation is to build new solutions of the
string theory field equations which could serve as models of the pre-Big
Bang, before the initial singularity, in the framework of Veneziano's pre-Big
Bang cosmology. 
\vspace*{3mm} \\
A credible implementation of this scenario requires the use of realistic
models for the primordial Universe, i.e. anisotropic and inhomogeneous
models. Some models among those considered in section 3 could be analyzed in
more detail: in particular, the geometrical and physical properties of the
exact dual solutions explicitly built could be studied in view of examining
the presence of a singularity and the existence of a pre-Big Bang
inflationary phase. 
\vspace*{3mm} \\
Another important point to consider is : ``Could there remain in the present
Universe some relics from an eventually pre-Big Bang phase that could involve
some observational evidence ?".
\vspace*{3mm} \\
In the same way, the problem of the possible avoidance of the black hole
singularity could be tackled using e.g. a black hole solution joined to its
dual solution (see e.g. \S\, 3.7) before reaching the singularity. This leaves
the door open to a series of speculative idea: is it possible, in this
framework, for the dual solution of a black hole to be an open door to 
another Universe ? 
In any case, this remains a challenge for our imagination... \\

\noindent \Large \bf  Acknowledgements \it \\
\normalsize

This work was supported in part by Belgian Interuniversity Attraction Pole
P4/O5 as well as by a grant from ``Fonds National de la Recherche
Scientifique".\\

\end{document}